\documentclass[%
 reprint,
superscriptaddress,
longbibliography,
 amsmath,amssymb,
 nobalancelastpage,
prb,
]{revtex4-2}
\usepackage[romanian,english]{babel}
\usepackage{combelow}
\usepackage{color}
\usepackage{multirow}

\useshorthands{'}
\defineshorthand{'t}{\cb{t}}
\defineshorthand{'S}{\cb{S}}
\defineshorthand{'T}{\cb{T}}

\usepackage{graphicx}
\usepackage{dcolumn}

\usepackage{bm}
\usepackage{amsmath}

\begin{document}
\preprint{APS/123-QED}

\title{Gradual emergence of superconductivity in underdoped La$_{2-x}$Sr$_x$CuO$_4$}

\author{Ana-Elena 'Tu'tueanu}
 \affiliation{Nanoscience Center, Niels Bohr Institute, University of Copenhagen, 2100 Copenhagen, Denmark} 
 \affiliation{Institut Laue-Langevin, 38042 Grenoble, France}
 
 \author{Machteld E. Kamminga }
\affiliation{Nanoscience Center, Niels Bohr Institute, University of Copenhagen, 2100 Copenhagen, Denmark}

 \author{Tim B. Tejsner}
\affiliation{Nanoscience Center, Niels Bohr Institute, University of Copenhagen, 2100 Copenhagen, Denmark}
\affiliation{Institut Laue-Langevin, 38042 Grenoble, France}

\author{Henrik Jacobsen}
\affiliation{Nanoscience Center, Niels Bohr Institute, University of Copenhagen, 2100 Copenhagen, Denmark}

\affiliation{Paul Scherrer Institute, Laboratory for Neutron Scattering and Imaging, 5232 Villigen, Switzerland}

\author{Henriette W. Hansen}
\affiliation{Nanoscience Center, Niels Bohr Institute, University of Copenhagen, 2100 Copenhagen, Denmark}

\author{Monica-Elisabeta L\u{a}c\u{a}tu\cb{s}u}
\affiliation{Nanoscience Center, Niels Bohr Institute, University of Copenhagen, 2100 Copenhagen, Denmark}
\affiliation{Paul Scherrer Institute, Laboratory for Neutron Scattering and Imaging, 5232 Villigen, Switzerland}

\author{Jacob Baas}
\affiliation{Zernike Institute for Advanced Materials, University of Groningen, 9747 AG Groningen, The Netherlands} 

\author{Kira L. Eliasen}
\affiliation{Nanoscience Center, Niels Bohr Institute, University of Copenhagen, 2100 Copenhagen, Denmark}
\affiliation{Glass and Time, IMFUFA, Department of Science and Environment, Roskilde University, 4000 Roskilde, Denmark}

\author{Jean-Claude Grivel}
\affiliation{Department of Energy Conversion and Storage, Technical University of Denmark, 2800 Kongens Lyngby, Denmark}

\author{Yasmine Sassa}
\affiliation{Department of Physics, Chalmers University of Technology, 412 96 Göteborg, Sweden}

\author{Niels Bech Christensen}
\affiliation{Department of Physics, Technical University of Denmark, 2800 Kgs. Lyngby, Denmark}

\author{Paul Steffens}
\affiliation{Institut Laue-Langevin, 38042 Grenoble, France}%

\author{Martin Boehm}
\affiliation{Institut Laue-Langevin, 38042 Grenoble, France}%

\author{Andrea Piovano}
\affiliation{Institut Laue-Langevin, 38042 Grenoble, France}%

\author{Kim Lefmann}%
\affiliation{%
Nanoscience Center, Niels Bohr Institute, University of Copenhagen, 2100 Copenhagen, Denmark}%

\author{Astrid T. R\o mer}
\email{asr@dfm.dk}
\affiliation{Nanoscience Center, Niels Bohr Institute, University of Copenhagen, 2100 Copenhagen, Denmark}
\affiliation{Danish Fundamental Metrology, Kogle All\'e 5, 2970 H{\o}rsholm, Denmark}

\date{\today}

\begin{abstract}
We present triple-axis neutron scattering studies of low-energy magnetic fluctuations in strongly underdoped La$_{2-x}$Sr$_{x}$CuO$_{4}$ with $x=0.05$, $0.06$ and $0.07$, providing quantitative evidence for a direct competition between these fluctuations and superconductivity. At dopings $x=0.06$ and $x=0.07$, three-dimensional superconductivity is found, while only a very weak signature of two-dimensional superconductivity residing in the CuO$_2$ planes is detectable for $x=0.05$. We find a surprising suppression of the low-energy fluctuations by an external magnetic field at all three dopings. This implies that the response of two-dimensional superconductivity to a magnetic field is similar to that of a bulk superconductor. Our results provide direct evidence of a very gradual onset of superconductivity in cuprates.

\end{abstract}

\maketitle

The emergence of unconventional superconductivity is closely connected to other electronic ordering phenomena, such as spin- and charge-density modulations.~\cite{Tranquada1995} In the simplest family of cuprate superconductors, antiferromagnetic order of the insulating parent compound La$_2$CuO$_4$ is suppressed by doping, and superconductivity emerges upon further doping.~\cite{tranquadastripe,Vojta2009a} In the low-doping regime spin- and charge-stripe order coexists with superconductivity and here, superconducting critical temperatures remain modest with $T_c \leq 20$ K. In general, low-energy magnetic fluctuations appear to be antagonists to superconductivity~\cite{Millis1988} and the opening of a gap in the magnetic energy spectrum is observed in cuprate superconductors only when these become doped to optimal critical temperature. For La$_{2-x}$Sr$_x$CuO$_4$ (LSCO) this happens around a doping of $x\gtrsim 0.14$.~\cite{Chang2009,lake1999gap} 

An external magnetic field applied to optimally doped LSCO generates in-gap states \cite{lake2001spins} and eventually
completely closes the superconducting gap.~\cite{Chang2009} 
In contrast, moderately underdoped LSCO ($x \sim 10$\%) shows no spin gap and no field effect on the low-temperature, low-energy excitations. \cite{chang2007magnetic} 
While there is consensus that optimally doped cuprates exhibit homogeneous $d$-wave superconductivity,~\cite{Harlingen95} 
the nature of superconductivity within the stripe-ordered phase on the underdoped side of the phase diagram is still being debated; is superconductivity phase-separated from stripe-ordered regions or do these coexist? Coexistence of stripes and
superconducting order in which the latter is spatially modulated to accommodate the spin stripe is foundin  La$_{2-x}$Ba$_{x}$CuO$_{4}$ (LBCO.)~\cite{tranquada2008evidence,Lozano21,tranquadastripe,XuTranquada2014,Christensen2016} This phase, dubbed the Pair-Density-Wave (PDW) phase, was also proposed on theoretical grounds in order to describe multiple intertwined orders in the low-doping regime of the cuprates.~\cite{Berg_2009,AgterbergPDW}
\begin{figure}[b!]
    \centering
    \includegraphics[width=0.8\columnwidth]{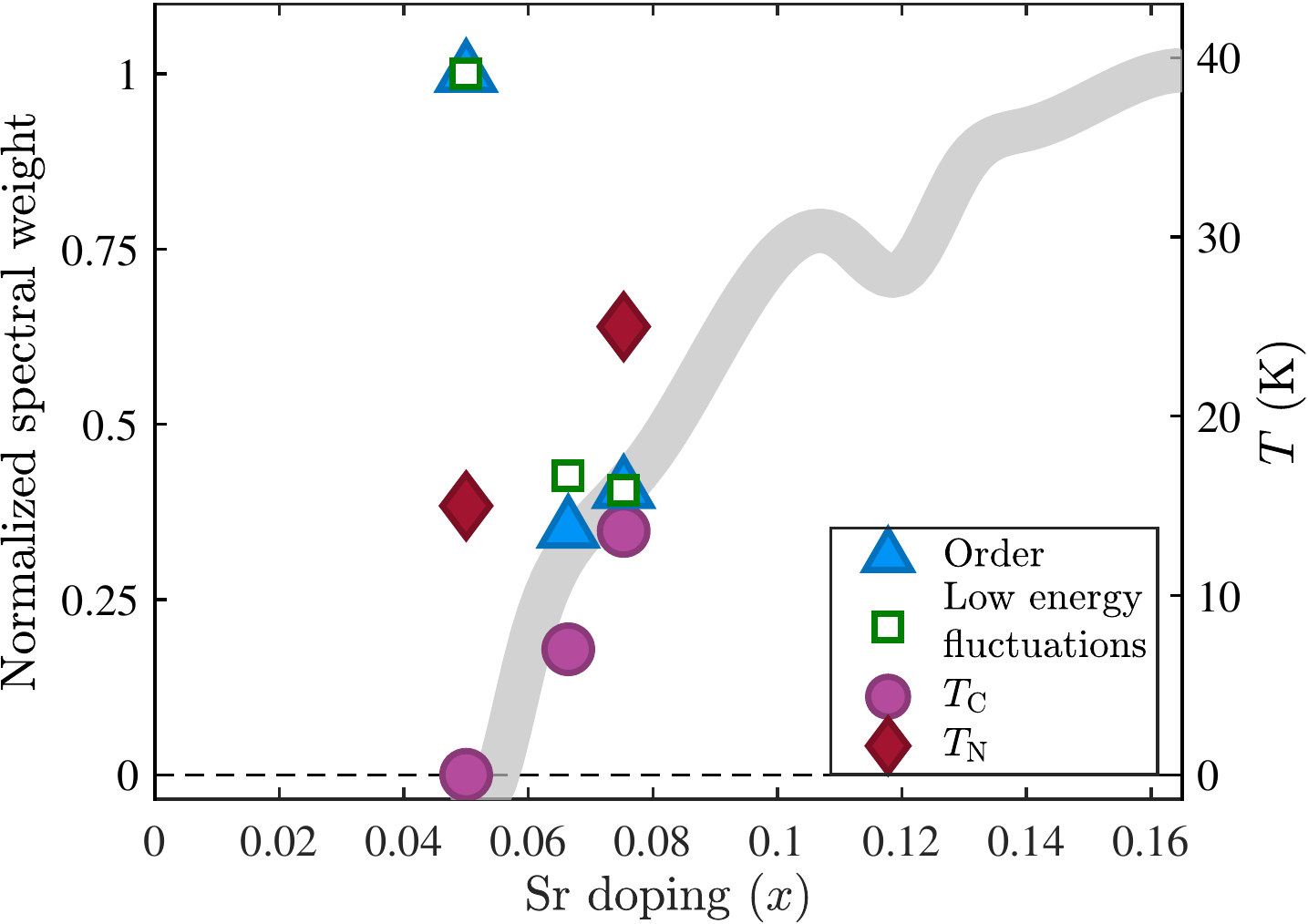}
 \caption{The evolution of magnetic order, low-energy fluctuations and superconducting $T_c$ as a function of doping. $T_c$ of our crystals (determined from SQUID measurements) is shown in magenta and the change in magnetic order and low-energy fluctuations ($\hbar \omega=0.8$ meV) quantified by the neutron scattering spectral weight is shown by blue triangles and green squares, respectively. The gray line displays the evolution of $T_c$ as determined in Refs.~\onlinecite{lake2001spins,Lake2002,chang2007magnetic,Kofu09}. 
}
    \label{fig:phdia}
\end{figure}
One key signature of the PDW phase is that two-dimensional superconductivity sets in at temperatures exceeding the bulk $T_c$. This can be seen by the onset of diamagnetism at $T>T_c$ for magnetic fields applied perpendicular to the CuO$_2$ planes.~\cite{tranquada2008evidence} For magnetic fields parallel to the planes, no diamagnetism is observed, which is interpreted as an interplanar frustration of the Josephson coupling preventing three-dimensional $d$-wave superconductivity from developing. An additional phase-sensitive confirmation of the PDW phase in LBCO was recently presented in Ref.~\onlinecite{Lozano21}.
Besides the growing evidence of a PDW phase in LBCO, scanning tunneling microscopy on the related material, Bi$_2$Sr$_2$CaCu$_2$O$_{8+x}$, found direct evidence of a modulated superconducting order parameter.\cite{Hamidian2016}

It is tempting to ask if spatially modulated superconductivity is a universal player encountered on the route to superconductivity in the cuprate family. To approach this  question, we focus on LSCO in the underdoped region near the onset of superconductivity ($x_c=0.055$) \cite{Takagi1989}. 

We first show that for $x<x_c$, this material displays weak two-dimensional superconducting correlations, a precursor to bulk superconductivity occurring at slightly larger doping values.

Next, we study the low-energy magnetic fluctuations and magnetic order using neutron scattering. Fig.~\ref{fig:phdia} shows the elastic and low-energy inelastic (0.8 meV) spin correlation functions as a function of doping, together with the superconducting transition temperatures. It is clear that both neutron signals drastically decrease at the onset of superconductivity. This is direct evidence of competition between superconductivity and low-energy fluctuations.

Finally, we show that the response to an external magnetic field does not depend on whether bulk superconductivity is established in the material or not. Our results provide evidence of a very gradual emergence -- rather than an abrupt onset -- of superconductivity in LSCO and point towards the universality of the PDW picture in the cuprates.

\begin{figure}[t!]
\centering
\includegraphics[width=0.5\textwidth]{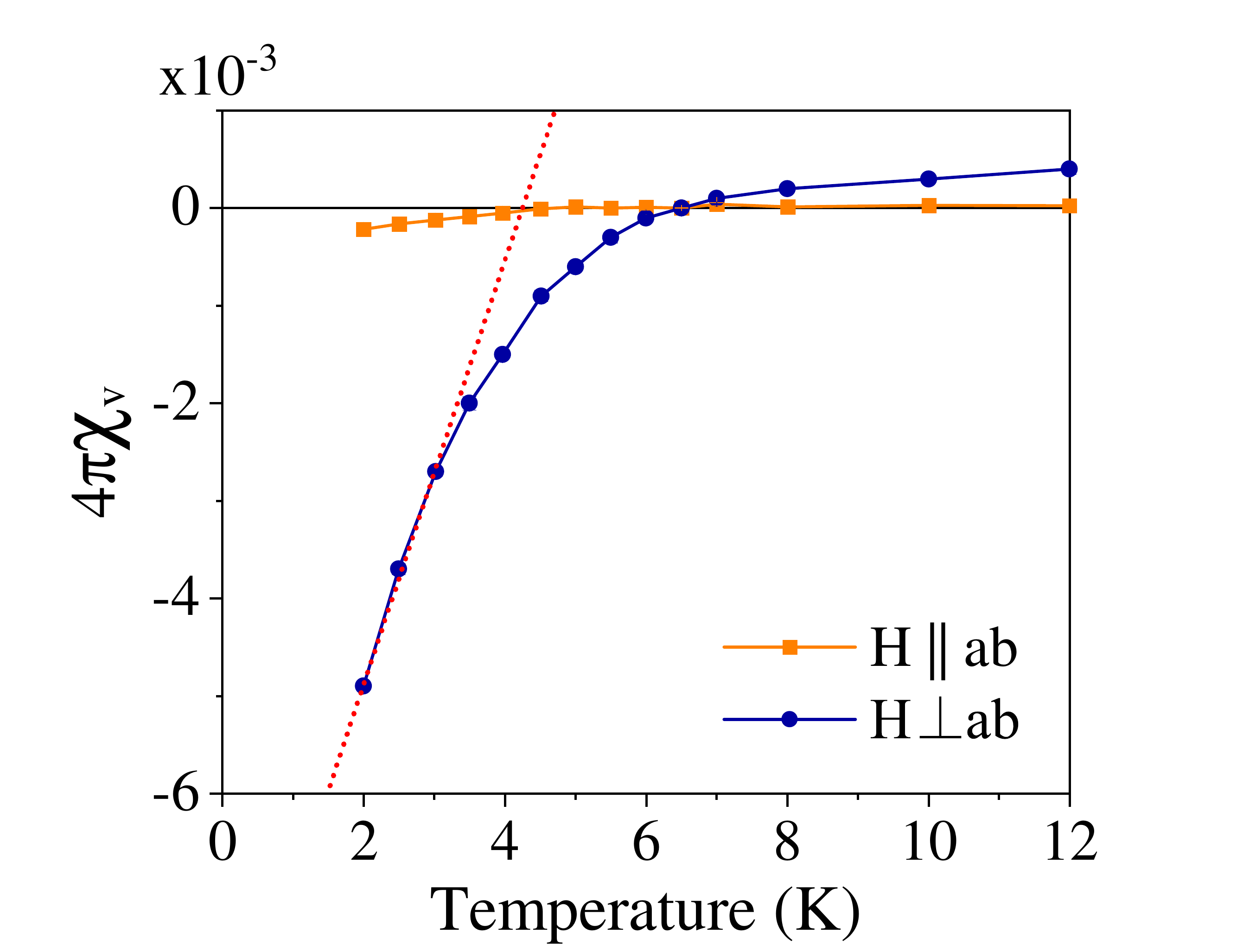}
\caption{ Magnetic volume susceptibility measured in LSCO5 with a small magnetic field ($H= 2$ mT) applied parallel and perpendicular to the $(a,b)$ plane. These measurements were performed while heating, preceded by zero field cooling (ZFC). A weak diamagnetic response sets in at the characteristic temperature of 4.2 $\pm$ 0.3 K displayed by the dotted red line.}
\label{fig:suscepLSCO5}
\end{figure}

 
We investigated three LSCO crystals with nominal doping of $x = 0.05$ (LSCO5), $0.06$ (LSCO6) and $0.07$ (LSCO7). The crystals were grown by the travelling-solvent floating-zone method,\cite{Tanaka89} 
and the precise doping was inferred 
from the transition temperature between the high-temperature tetragonal and the low-temperature orthorhombic phases,\cite{Fujita2002} yielding $x = 0.050(2)$, $0.0664(6)$, and $0.0753(3)$, for LSCO5, LSCO6 and LSCO7, respectively. In the crystal with the lowest dopeing, LSCO5, susceptibility measurements in a magnetic field applied perpendicular to the CuO$_2$ planes find a weak but clear diamagnetic response (blue data points in Fig.~\ref{fig:suscepLSCO5}). 
We interpret this as formation of screening currents within the CuO$_2$ plane in small parts of the sample at temperatures below 4.2 $\pm$ 0.3 K.
On the other hand, superconducting correlations between the planes are strongly suppressed, and only an extremely weak response is found for fields applied parallel to the CuO$_2$ planes (orange data points in Fig.~\ref{fig:suscepLSCO5}). This response is most likely simply caused by a minor misalignment.  More details are given in the SM.\cite{supplemental} 
The superconducting transition temperatures of LSCO6 ($T_c=7.0 \pm 0.4$ K) and LSCO7 ($T_c=13.6 \pm 0.2$ K) are determined from susceptibility measurements on small, crushed
pieces of the corresponding large single crystals used for the neutron studies presented throughout this paper, 
see SM for more details.\cite{supplemental} For both samples, we find that the diamagnetic response is far from that of a perfect diamagnet at $T=2$ K, indicating that a sizeable fraction of the samples remains non-superconducting at this temperature. The increase of $T_c$ upon doping is shown by the magenta points and the gray curve in Fig.~\ref{fig:phdia}. Critical temperatures at higher doping values are inferred from literature.~\cite{lake2001spins,Lake2002,chang2007magnetic,Kofu09} We note that the width of the line is a reflection of the spread in the reported values of $T_c$.

\begin{figure*}[t!]
\centering\includegraphics[width=\textwidth]{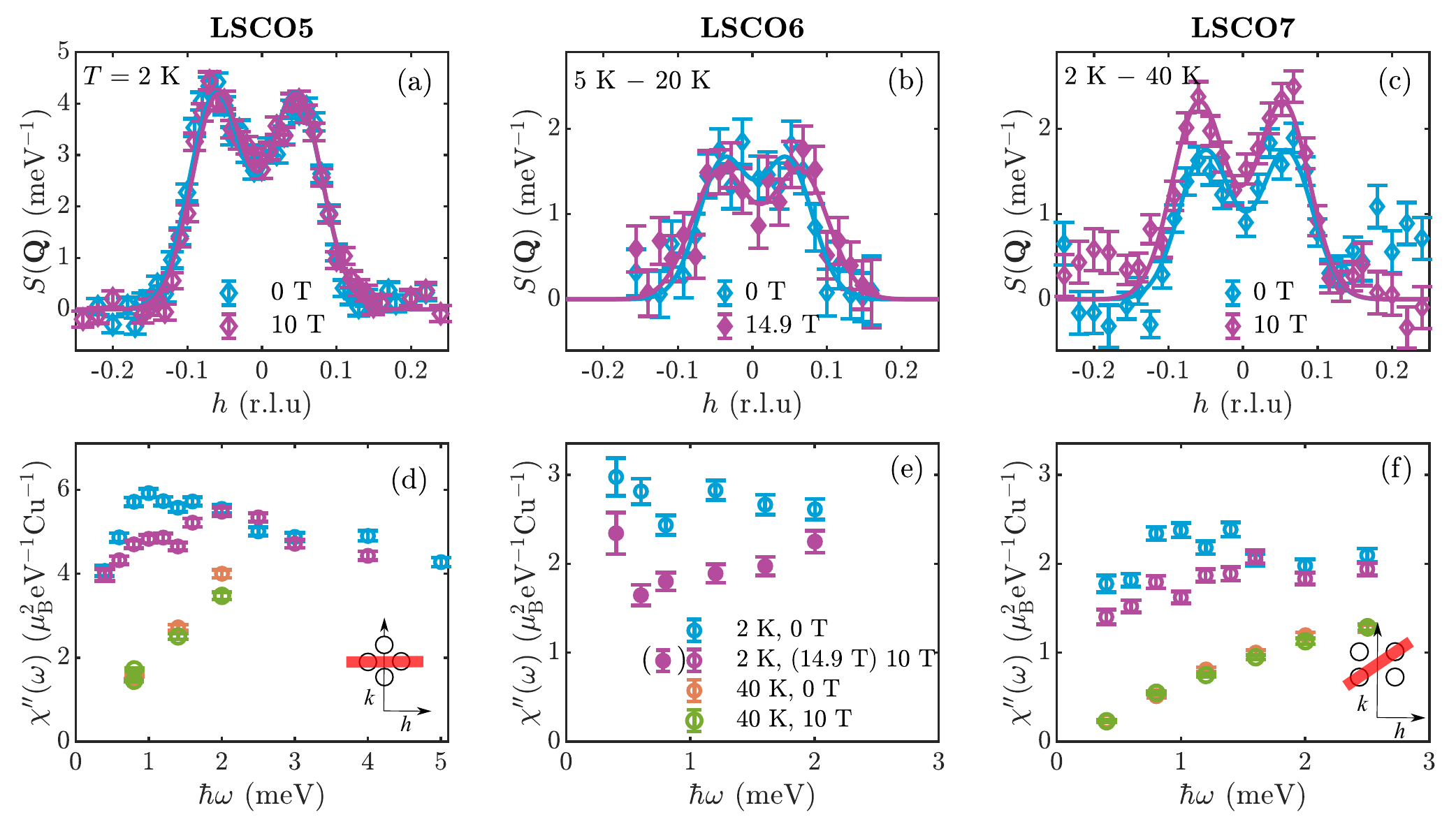}
\caption{\textbf{(a)-(c)} Spin correlation $S({\bf{Q}})$ of background subtracted elastic signals with (purple) and without (blue) applied magnetic field of the samples LSCO5, LSCO6 and LSCO7, respectively. 
The solid lines are Gaussian fits as explained in the text. \textbf{(d)-(f)} Brillouin Zone averaged dynamic susceptibility $\chi^{\prime\prime} (\omega)$ measured at $2$~K at zero field (blue circles) and with an applied magnetic field of $10$~T (purple circles), and at $40$~K at zero field (orange circles) and with an applied magnetic field of $10$~T (green circles). For the LSCO6 sample, the magnetic field was $14.9$~T (represented by the filled purple circles). The inset in (d) illustrates the scan direction used for the LSCO5 sample, while the inset in (f) shows the scan direction for the LSCO6 and LSCO7 samples. 
}
\label{fig:Figure1}
\end{figure*}

To study the incommensurate magnetic order and fluctuations, we performed several experiments at the cold neutron triple-axis spectrometer ThALES~\cite{boehm2015thales} at Institut Laue-Langevin (ILL). \cite{ThALES_data1, ThALES_data2, ThALES_data3, ThALES_data4} The samples were placed in $10$~T (LSCO5 and LSCO7) and $15$~T (LSCO6) vertical field cryomagnets to vary the temperature between 2 and 45 K, and apply high magnetic fields. Details of the experimental setup can be found in the SM.\cite{supplemental}
With the $c$ axis vertical, scattering wave vectors ${\bf{Q}} = (h, k, 0)$ were accessible in the horizontal scattering plane. Wave vectors are expressed in units of $(2\pi/a,2\pi/b,2\pi/c)$ with the measured low-temperature lattice parameters $a\approx 5.38$ \AA{}, $b\approx 5.39$ \AA{}, and $c\approx 13.2$ \AA{} in the orthorhombic setting (space group $Bmab$, $I4/mmm$ above $T_N$), depending on doping. 
At doping values below $x\lesssim 0.06$ incommensurate magnetic order and low-energy fluctuations can be observed at ${\bf{Q}}=(1\pm\delta,0,0),(0,1\pm\delta,0)$. 
At higher doping values, above the onset of bulk superconductivity, a $45^{\circ}$ rotation of the magnetic ordering vector occurs, \textit{i.e.}
${\bf{Q}}=(1\pm\delta,\pm\delta,0),(\pm\delta,1\pm\delta,0)$,\cite{Fujita2002} see the insets in Fig.~\ref{fig:Figure1}(d,f) and SM for details.~\cite{supplemental}

Despite $\delta$ being very small in the doping regime of this work, we were able to distinguish the incommensurate peaks in all our samples, see Figs.~\ref{fig:Figure1}(a-c). 
All data has been fitted with the same double Gaussian model with equal amplitudes and widths for the two peaks. The data are normalized to absolute units using the intensity of the acoustic phonons following the procedure proposed by G. Xu {\em et al}.~\cite{xu2013absolute} This allows for a direct quantitative comparison between generalized magnetic susceptibilities of samples of different doping. Note that the signal from the LSCO5 sample  is roughly twice as large as in the crystals of higher doping.
In regard to the low-energy magnetic fluctuations, we likewise observe that  LSCO5 shows a generalized spin susceptibility which is approximately twice as large as in LSCO6 and LSCO7, see Fig.~\ref{fig:Figure1}(d-f).
No gap opening occurs down to the lowest energy transfers measured, $\hbar\omega=0.4$~meV in any of the crystals. This behavior is reminiscent of other cuprate materials that display static order, see for example Refs.~\onlinecite{chang2007magnetic,Matsuda08,Kofu09,Haug09,XuTranquada2014,Jacobsen2015,romer2013,Jacobsen18,tutueanu2021}.

To further investigate the magnetic behavior in all three crystals, we applied a magnetic field perpendicular to the CuO$_2$ planes. This causes a clear enhancement of the static signal in LSCO7, see Figs.~\ref{fig:Figure1}(c), while no discernible effect is seen in the static magnetic signal in  LSCO5 and LSCO6 as shown in Fig.~\ref{fig:Figure1}(a,b). This leads to the conclusion that the field effect in the elastic channel is largest for the crystal that displays the largest superconducting volume fraction. 
On the other hand, the field effect in the {\it inelastic} channel is very similar in all three samples, where we observe  a clear decrease in the dynamic susceptibility, 
$\chi''(\omega)$,  for $\hbar \omega <2$ meV, see Figs.~\ref{fig:Figure1} (d-f). This suppression is present at base temperature, whereas the low-energy susceptibility is field-independent at $T=40$ K.

This leaves us with the interesting observation that the low-energy response is similar in all samples, albeit different in overall magnitude, regardless of whether a coherent superconducting phase is present or not. This is surprising, given the clear difference in the magnetic field effect in the static signal of the different samples.  In LSCO7, the magnetic field suppression of the inelastic signal goes along with an enhancement of the static signal, see Fig.~\ref{fig:Figure1}(c). Such behavior can generally be interpreted as the slowing-down of spin fluctuations inside vortices of the CuO$_2$ planes. Intriguingly, our static measurements in LSCO5 or LSCO6 show no such effects. The vortex picture is thus insufficient to explain the data in the case of LSCO5 and LSCO6, but we hypothesize that the formation of two-dimensional superconducting islands could play a role in the field suppression of the low-energy fluctuations.

For LSCO5, we have performed 
additional experiments to verify that no spectral weight transfer towards higher energies up to $50$~meV takes place  (see SM \cite{supplemental}). 
Notably, the onset of the field effect in the inelastic channel in LSCO5 occurs at T $\geq15$~K, see SM. \cite{supplemental}

\begin{figure}[t!]
\centering
\includegraphics[width=0.8\columnwidth]{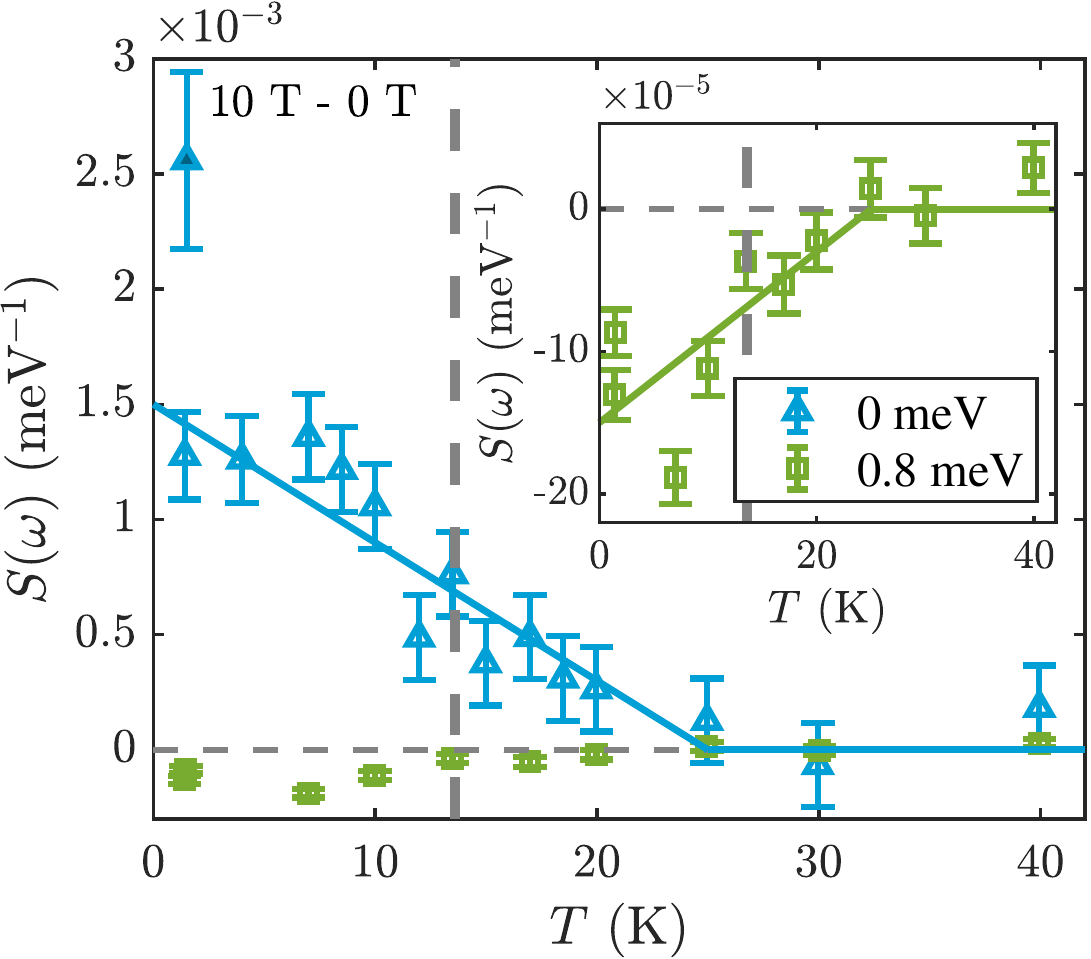}
\caption{Temperature dependence of the field effect on magnetic order and fluctuations ($\hbar \omega=0.8$ meV) in LSCO7 at $Q_\mathrm{IC}=(0.93,0.07,0)$.
The data is shown as point-by-point subtraction of the Brillouin Zone averaged spin correlation function at zero field from that at $10$~T. The insert shows the 0.8 meV data on a smaller scale. The vertical dashed lines denote $T_\textnormal{c}$, and solid lines are guides to the eye.}
\label{fig:field temp commparison el inel LSCO7}
\end{figure}

For LSCO7, the temperature dependence of the magnetic field effect on the elastic and inelastic signals is shown in Fig.~\ref{fig:field temp commparison el inel LSCO7}. This reveals that the onset of the field enhancement in the elastic channel occurs at $T_\text{field} \geq 25$~K, well above $T_\textnormal{c}$. The field-suppression of the fluctuations at $\hbar \omega = 0.8$~meV occurs at the same temperature. This indicates a spectral weight shift from the low-energy magnetic excitations to the elastic channel.

We interpret the fact that $T_\text{field}>T_\textnormal{c}$ to be a result of two-dimensional superconducting correlations being present in the material already at temperatures above $T_\textnormal{c}$. Indeed, in the $x=0.07$ doped crystal investigated in Ref.~\onlinecite{Jacobsen2015}, the diamagnetic response for in-plane fields occurs prior to the onset of three-dimensional superconductivity. This interpretation is consistent with the fact that two-dimensional superconducting correlations appear to be sufficient to trigger a spectral weight shift of magnetic intensity in LSCO5, that also exhibits weak 2D superconductivity as demonstrated in Fig.~\ref{fig:suscepLSCO5}. Our findings point towards a gradual emergence of superconductivity in the underdoped part of the phase diagram both as a function of cooling, but also as a function of doping. In both cases,  two-dimensional superconducting correlations appear to be the common precursor. 

Interestingly, a magnetic field applied perpendicular to the CuO$_2$ planes suppresses the low-energy fluctuations in all three underdoped samples, in contrast to what is seen in optimally doped LSCO.  We speculate that this suppression is related to the presence of two-dimensional superconducting correlations even at the lowest doping values and at elevated temperatures.

Lastly, we discuss our findings in relation to other members of the cuprate family. The oxygen-doped cuprate compound La$_2$CuO$_{4+y}$ with much higher $T_\textnormal{c}=42$~K, also displays an ungapped spectrum\cite{wells1997,Lee1999} and a weak magnetic field suppression of the low-energy fluctuations.\cite{tutueanu2021}
A similar behavior was also found in underdoped YBa$_2$Cu$_3$O$_{6.45}$ with $T_c=35$ K.~\cite{Haug09} 
Common for the above-mentioned systems as well as for LBCO~\cite{Wen08} is  that  low-energy fluctuations become prominent at low temperatures ($T<T_\textnormal{c}$) and suppressed by an applied magnetic field.
Surprisingly, this common behavior is observed for cuprates at vastly different positions in the phase diagram.

In conclusion, we find that features of PDW show up throughout the underdoped part of the phase diagram of LSCO.
We propose that the emergence of superconductivity involves both  PDW, and gradually larger uniform $d$-wave superconducting regions which are phase-seperated from the stripe-ordered regions. The ratio between the two SC phases can be tuned by varying the strontium doping of the samples, while an applied magnetic field perpendicular to the CuO$_2$ planes enhances magnetic order suppressing at the same time superconductivity irrespective of the type of superconductivity. 
We highlight the observation that a similar response to a magnetic field is observed across the cuprate family, and that the gradual emergence of superconductivity involving the development of two-dimensional correlations prior to the homogeneous $d$-wave phase could likewise be universal across the cuprates.

\section*{Acknowledgements}
We are grateful for the access to neutron beamtime at the ILL neutron facility. AET was supported through the ILL Ph.D. program. ATR and HJ acknowledge support from the Carlsberg Foundation. MEK was supported by the EU MSCA program through grant number 838926. We thank Jesper Bendix for providing magnetometry measurement time. The project was supported by the Danish National Committee for Research Infrastructure through DANSCATT and the ESS-Lighthouse Q-MAT. 

\bibliography{LSCO_magnetism_1.bib}

\end{document}


\preprint{APS/revtex4-2}

\title{Supplemental Material to: Gradual emergence of superconductivity in underdoped La$_{2-x}$Sr$_x$CuO$_4$}

\author{Ana-Elena 'Tu'tueanu}
 \affiliation{Nanoscience Center, Niels Bohr Institute, University of Copenhagen, 2100 Copenhagen, Denmark} 
 \affiliation{Institut Laue-Langevin, 38042 Grenoble, France}
 \email{tutueanu@ill.fr}
 
 \author{Tim B. Tejsner}
\affiliation{Nanoscience Center, Niels Bohr Institute, University of Copenhagen, 2100 Copenhagen, Denmark}
\affiliation{Institut Laue-Langevin, 38042 Grenoble, France}

\author{Machteld E. Kamminga }
\affiliation{Nanoscience Center, Niels Bohr Institute, University of Copenhagen, 2100 Copenhagen, Denmark}

\author{Henrik Jacobsen}
\affiliation{Nanoscience Center, Niels Bohr Institute, University of Copenhagen, 2100 Copenhagen, Denmark}
\affiliation{Paul Scherrer Institute, Laboratory for Neutron Scattering and Imaging, 5232 Villigen, Switzerland}

\author{Henriette W. Hansen}
\affiliation{Nanoscience Center, Niels Bohr Institute, University of Copenhagen, 2100 Copenhagen, Denmark}

\author{Monica-Elisabeta L\u{a}c\u{a}tu\cb{s}u}
\affiliation{Nanoscience Center, Niels Bohr Institute, University of Copenhagen, 2100 Copenhagen, Denmark}
\affiliation{Paul Scherrer Institute, Laboratory for Neutron Scattering and Imaging, 5232 Villigen, Switzerland}

\author{Jacob Baas}
\affiliation{Zernike Institute for Advanced Materials, University of Groningen, 9747 AG Groningen, The Netherlands} 

\author{Kira L. Eliasen}
\affiliation{Nanoscience Center, Niels Bohr Institute, University of Copenhagen, 2100 Copenhagen, Denmark}
\affiliation{Glass and Time, IMFUFA, Department of Science and Environment, Roskilde University, 4000 Roskilde, Denmark}

\author{Jean-Claude Grivel}
\affiliation{Department of Energy Conversion, Technical University of Denmark, 4000 Roskilde, Denmark}

\author{Yasmine Sassa}
\affiliation{Department of Physics, Chalmers University of Technology, 412 96 Göteborg, Sweden}

\author{Niels Bech Christensen}
\affiliation{Department of Physics, Technical University of Denmark, 2800 Kgs. Lyngby, Denmark}

\author{Paul Steffens}
\affiliation{Institut Laue-Langevin, 38042 Grenoble, France}%

\author{Martin Boehm}
\affiliation{Institut Laue-Langevin, 38042 Grenoble, France}%

\author{Andrea Piovano}
\affiliation{Institut Laue-Langevin, 38042 Grenoble, France}%

\author{Kim Lefmann}%
\affiliation{%
Nanoscience Center, Niels Bohr Institute, University of Copenhagen, 2100 Copenhagen, Denmark}%

\author{Astrid Tranum R\o mer}
\affiliation{Nanoscience Center, Niels Bohr Institute, University of Copenhagen, 2100 Copenhagen, Denmark}
\email{asr@dfm.dk}

\date{\today}

\pacs{74.72.-h,75.25.-j,75.40.Gb,78.70.Nx}
\begin{center}
 {\Large \textbf{Supplementary Material to: Gradual emergence of superconductivity in underdoped La$_{2-x}$Sr$_x$CuO$_4$}}

\hfill \\

{\normalsize Ana-Elena 'Tu'tueanu,$^{1, 2}$  Machteld E. Kamminga,$^{1}$ Tim B. Tejsner,$^{1, 2}$  Henrik Jacobsen,$^{1, 3}$ Henriette W. Hansen,$^{1}$ Monica-Elisabeta L\u{a}c\u{a}tu\cb{s}u,$^{1, 3}$ Jacob Baas,$^{4}$ Kira L. Eliasen,$^{1, 5}$ Jean-Claude Grivel,$^{6}$  Yasmine Sassa,$^{7}$ Niels Bech Christensen,$^{8}$   Paul Steffens,$^{2}$ Martin Boehm,$^{2}$ Andrea Piovano,$^{2}$ Kim Lefmann,$^{1}$ and Astrid T. Rømer,$^{1,9}$ }

\hfill \\

{\small $^1$Nanoscience Center, Niels Bohr Institute, University of Copenhagen, 2100 Copenhagen, Denmark
\par
$^2$Institut Laue-Langevin, 38042 Grenoble, France\par
$^3$Paul Scherrer Institute, Laboratory for Neutron Scattering and Imaging, 5232 Villigen, Switzerland\par
$^4$Zernike Insitute for Advanced Materials, University of Groningen, 9747 AG Groningen, The Netherlands\par
$^5$Glass and Time, IMFUFA, Department of Science and Environment, Roskilde University, 4000 Roskilde, Denmark\par
$^6$Department of Energy Conversion, Technical University of Denmark, 4000 Roskilde, Denmark\par
$^7$Department of Physics, Chalmers University of Technology, 412 96 Göteborg, Sweden\par
$^8$Department of Physics, Technical University of Denmark, 2800 Kgs. Lyngby, Denmark\par
$^9$Danish Fundamental Metrology, Kogle All\'e 5, 2970 H\o rsholm, Denmark}

\end{center}

\section{Magnetic susceptibility and superconducting critical temperature}
We determined the precise doping of the three samples by measuring the transition temperature between the high-temperature tetragonal and the low-temperature orthorhombic phases. The crystal masses, exact doping value and structural transition temperature for each crystal are given in Table~\ref{tab_collect}.

 \begin{table}[h!]
     \centering
     \begin{tabular}{ccccc}
 \hline
  \hline
Sample & Mass (g) & Doping ($x$) & $T_\textnormal{s}$ (K) & $T_\textnormal{c}$~(K)\\
 \hline 
 LSCO5 & 3.5 & $0.050(2)$& $416(5)$&  N/A   \\ 
 LSCO6 & 2.36 & $0.0664(6)$& $382(1)$& $7.0(4)$  \\ 
 LSCO7 & 3.4 & $0.0753(3)$& $362(1)$& $13.6(2)$  \\
 \hline
 \hline
     \end{tabular}
     \caption{Characterisation of the three \lsco{} samples of different strontium doping. 
     The structural transition temperature, $T_\textnormal{s}$ separating the high-temperature tetragonal (HTT) and the low-temperature orthorhombic (LTO) phases was determined through neutron diffraction experiments and
     the sample doping was deduced from $T_\textnormal{s}$~\cite{wakimoto2004neutron}.
     The superconducting critical temperature ($T_\textnormal{c}$) was determined by magnetic susceptibility measurements.}
     \label{tab_collect}
 \end{table}

\begin{figure}[h!]
\centering
\includegraphics[width=0.4\textwidth]{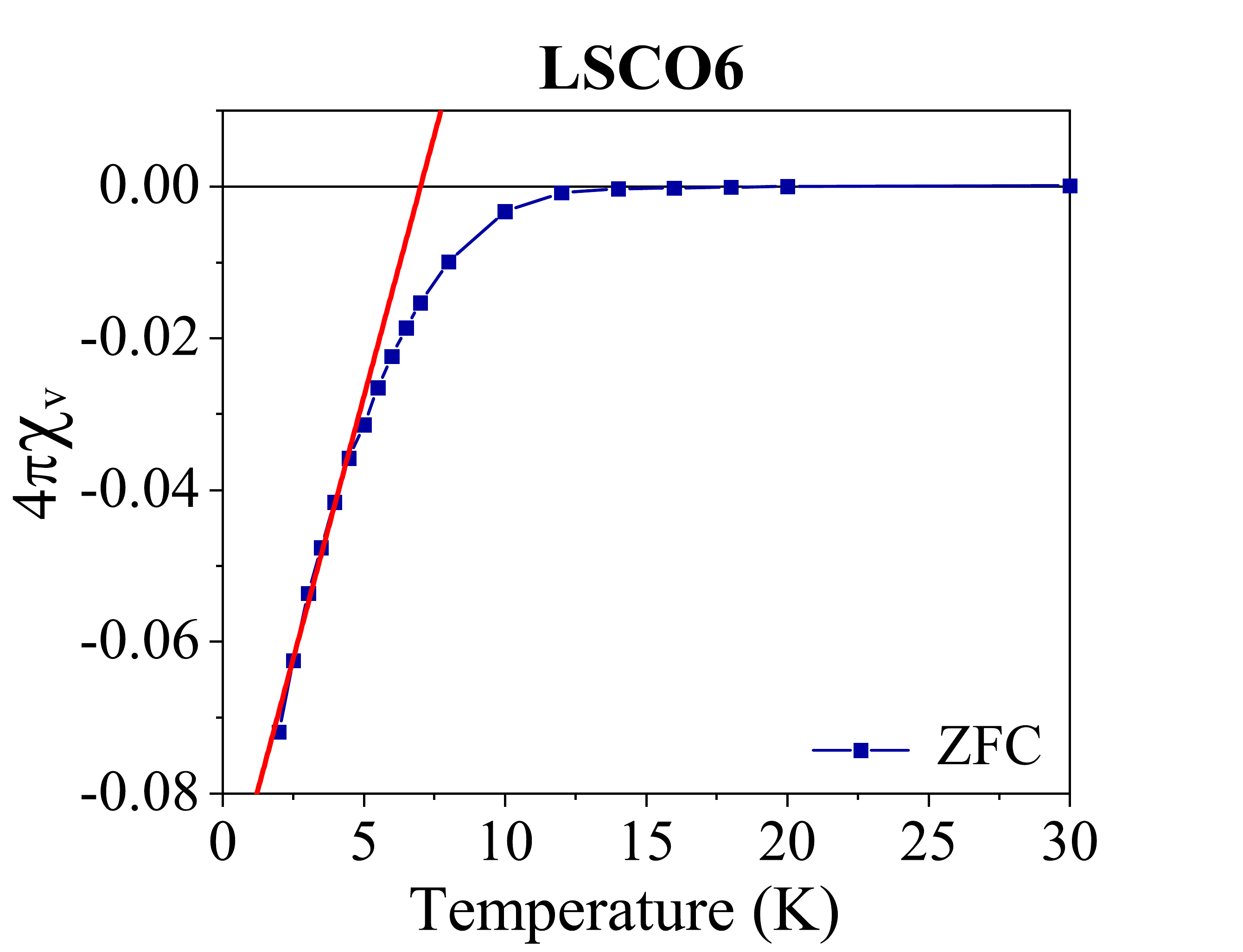}
\includegraphics[width=0.4\textwidth]{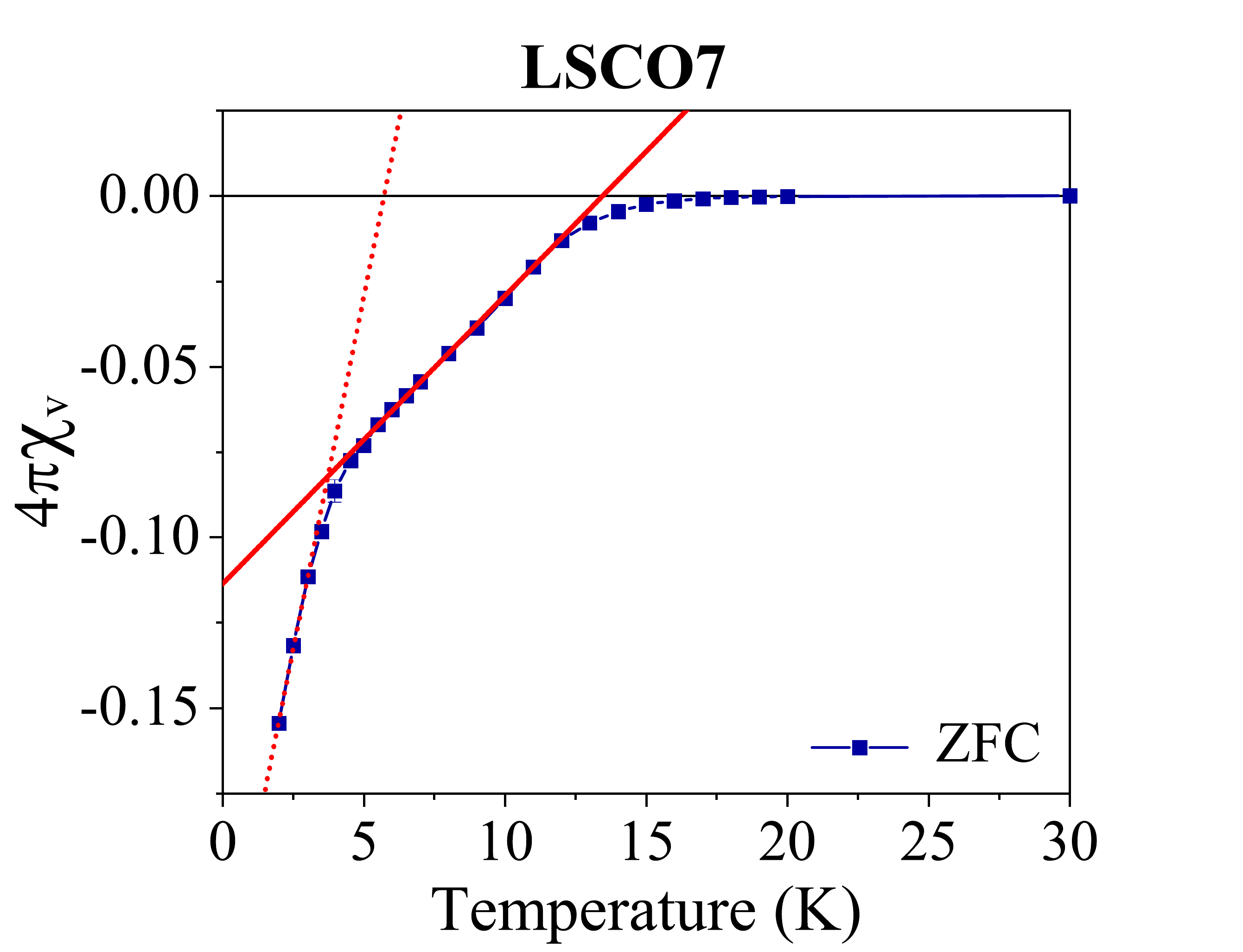}
\caption{ Magnetic (volume) susceptibility measured in powder samples of LSCO6 (left) and LSCO7 (right) with an applied magnetic field of 1 mT and 5 mT, respectively. These measurements were performed while heating, preceded by zero field cooling (ZFC). The red line corresponds to superconducting transition temperatures of 7.0 $\pm$ 0.4 K and 13.6 $\pm$ 0.2 K, for the LSCO6 and LSCO7 samples, respectively. The LSCO7 data has a second transition temperature of 5.7 $\pm$ 0.4 K, as denoted by the dotted red line.}
\label{fig:suscepLSCO6-7}
\end{figure}

Magnetic susceptibility measurements on the \lsco{} samples were conducted using a Quantum Design MPMS-XL superconducting quantum interference device (SQUID) magnetometer. All measurements on LSCO6 and LSCO7 were performed on small randomly oriented pieces of the corresponding large single crystals used for the neutron studies presented in this work. The measured signals were converted to volume susceptibility following Ref.~\onlinecite{tutueanu2021}. Fig.~\ref{fig:suscepLSCO6-7} shows the volume susceptibility for the LSCO6 and LSCO7 samples. The LSCO6 sample shows a clear superconducting phase transition at 7.0 $\pm$ 0.4 K, as determined by fitting the slope of the diagmagnetic response. The LSCO7 sample appears to have a phase inhomogeneity, resulting in two superconducting phase transitions: a low-temperature transition at 5.7 $\pm$ 0.4 K and a high-temperature transition at 13.6 $\pm$ 0.2 K. 
 We confirmed that the crystal was uniform by measuring very similar diamagnetic responses from crushed parts of both ends of the crystal.
 
For the LSCO5 sample, a diamond saw was used to cut a small sheet-like piece of 21.9 mg, with the $c$-axis oriented out of plane. The susceptibility measurements with the magnetic field oriented parallel and perpendicular to the $ab$-plane are shown in Fig. 2 of the main paper.


\section{Details on the neutron scattering experiments at ThALES (ILL)}
%
The instrument configuration that we used for the neutron scattering experiment on ThALES was the standard setting of the instrument, employing a velocity selector before the monochromator to suppress second-order contamination, and a Beryllium-filter before the analyser to reduce background. No collimation was used. The samples were placed in $10$~T (LSCO5 and LSCO7) and $15$~T (LSCO6) vertical field cryomagnets and slowly cooled with a $1$~K/min rate down to $100$~K and afterwards fast cooled. The outgoing wave vector, $k_\textnormal{f}=1.55$~\AA$^{-1}$, was kept constant across the different experiments since this value was found to be free of unwanted double scattering events~\cite{tutueanu2020multiple}. During data acquisition, the energy resolution was $\Delta E \sim 0.15$~meV (FWHM) in the elastic condition and increased up to $\sim 0.3$~meV for the inelastic scans depending on the energy transfer. In the standard focusing condition, applied in these experiments, the momentum resolution was $\Delta \QQQ \sim 0.02$~\AA$^{-1}$.

\section{Normalization to absolute units}
In this work, all data was normalized to absolute units using the method presented in Ref.~\onlinecite{xu2013absolute}. We outline the method here. For consistency, we always measured the phonon as a transverse scan at constant energy of 2 meV near (200) at 100 K.
The modified dynamic spin correlation function, $\tilde{S}(\QQQ,\omega)$, (modified by taking into account that neutrons are only sensitive to spin components perpendicular to $\QQQ$), is 
\begin{align}
    \tilde{S}(\QQQ,\omega) =\frac{13.77\, \text{barns}^{-1}  \tilde{I}(\QQQ,E)}{g^2|f(\QQQ)|^2 e^{-2W} Nk_fR_0}.
\end{align}
The imaginary part of the susceptibility, $\chi''(\QQQ,\omega)$, is given by
\begin{align}
    \chi''(\QQQ,\omega)=\frac{\pi}{2} \mu_B^2 (1-e^{-\hbar \omega/(k_BT)}) \frac{13.77\, \text{barns}^{-1} \tilde{I}(\QQQ,E)}{|f(\QQQ)|^2 e^{-2W} Nk_fR_0},
\end{align}
in units of $\mu_B^2$\,meV$^{-1}$, where $\mu_B$ is the Bohr magneton. In the above equations, $f(\QQQ)$ is the magnetic form factor of Cu, $e^{-2W}$ is the Debye-Waller factor, and $\tilde{I}(\QQQ,E)$ is the measured intensity divided by the monitor count and the energy of the magnetic excitation is given by  $\hbar \omega$. As usual, $k_B$ denotes the Boltzmanns constant and $g=2$ is  the Landé $g$ factor. Finally, $Nk_fR_0$ is a measure of the volume of the resolution function ($N$ is the number of unit cells). For phonons measured at constant energy it is

\begin{align}
    Nk_fR_0 = \frac{\int\tilde{I}(\QQQ,E) d\QQQ }{e^{-2W} |F_N(\GGG)|^2 \cos^2(\beta) \frac{m}{M} \frac{(\hbar \QQQ)^2}{2m} \frac{n}{\hbar \omega} \frac{1}{d\omega/dq} }.
\end{align}

$Nk_fR_0$ has units of meV/barn. $F_N$ is the structure factor of the acoustic phonon, which is the same as the Bragg structure factor at $\QQQ=\GGG(=(200)$ in our case), $\beta$ is the angle between $\QQQ$ and the polarization of the phonon, $m$ is the neutron mass, $M$ is the mass of all the atoms in the unit cell, $n$ is the Bose function for the phonon, and $d\omega /dq $ is the phonon velocity.  $\int\tilde{I}(\QQQ,E) d\QQQ$ is the integrated intensity of the phonon.

After converting the raw data to $\tilde{S}(\QQQ,\omega)$ or $\chi''(\QQQ,\omega)$ using the above equations, we averaged the signal over the Brillouin Zone. To this end, we first integrated the signal over the measured $h$ direction by fitting each peak to a Gaussian as described in the main text. To estimate the integral over the $k$ direction we assumed the signal was Gaussian along $k$ with the same width as along $h$. For the $l$ direction we assumed the signal was constant. Finally, we assumed that there are two peaks in the Brillouin Zone in the orthorhombic setting. 

\section{Orientation in reciprocal space of the magnetic signal}

\begin{figure}[h!]
\centering
\hspace*{-1cm}
\includegraphics[width=1.1\textwidth]{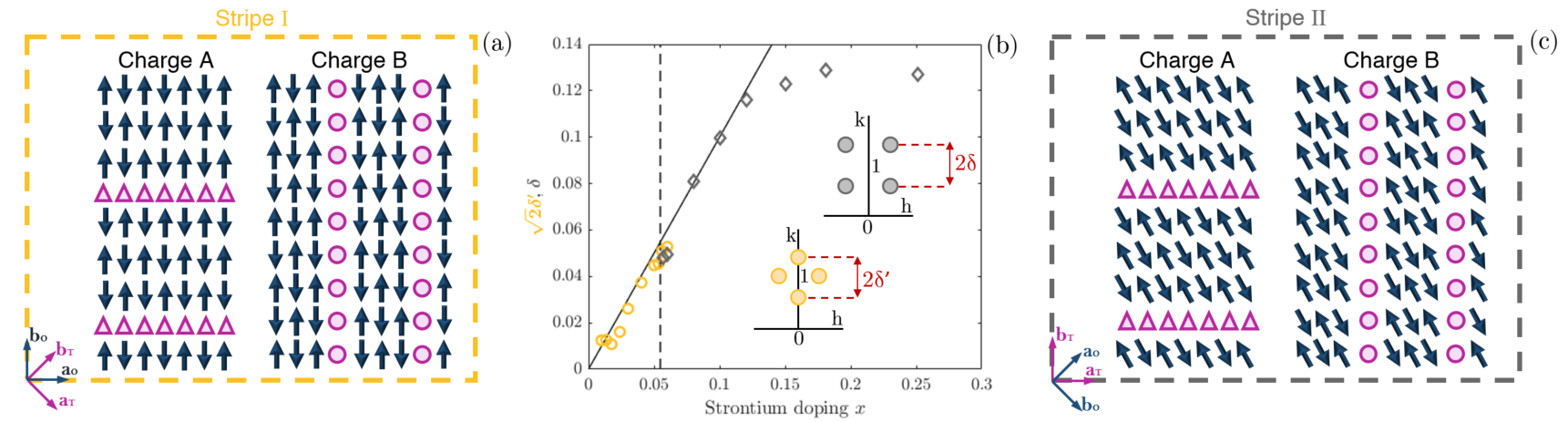}
\caption{\textbf{(a)} and \textbf{(c)} Graphical representation of stripes arrangement in both stripe I and stripe II phases respectively. In each case, the spin direction, along the orthorhombic $b$-axis, of the parent compound (LCO) is preserved~\cite{Lee1999a,vaknin1987antiferromagnetism}. The purple triangles and circles depict the doped charge rivers. The corresponding neutron scattering patterns are represented in the inset of subfigure (b), stripe I in yellow and stripe II in grey. The yellow points have been multiplied by a factor $\sqrt{2}$ in order to correct for the difference in symmetry so that, in both cases, the half difference in $Q$ between two neighbour IC peaks is plotted on the $y$-axis. Note that in both illustrations a $1/8$ periodicity is presented which is valid for $x\geqslant0.12$. \textbf{(b)} Doping dependence of the incommensurability over the entire phase diagram of LSCO. Yellow circles depict the highly underdoped and non-superconducting region in which stripes run along the orthorhombic $a/b$-axis (stripe I) while grey diamonds represent the superconducting region in which stripes run along the tetragonal $a/b$-axis (stripe II). Note that the insets showing the expected neutron pattern follow the orthorhombic notation, with $h/k$ reciprocal space directions superimposed on the orthorhombic $a/b$-axis. The solid line indicates the linear dependence $\delta = x$ while the dashed line marks the onset of superconductivity. Data reproduced from~\cite{vojta2009lattice}.}
\label{fig:stripes incommensurability}
\end{figure}

\begin{figure}[h!]
\centering
\includegraphics[width=0.43\textwidth]{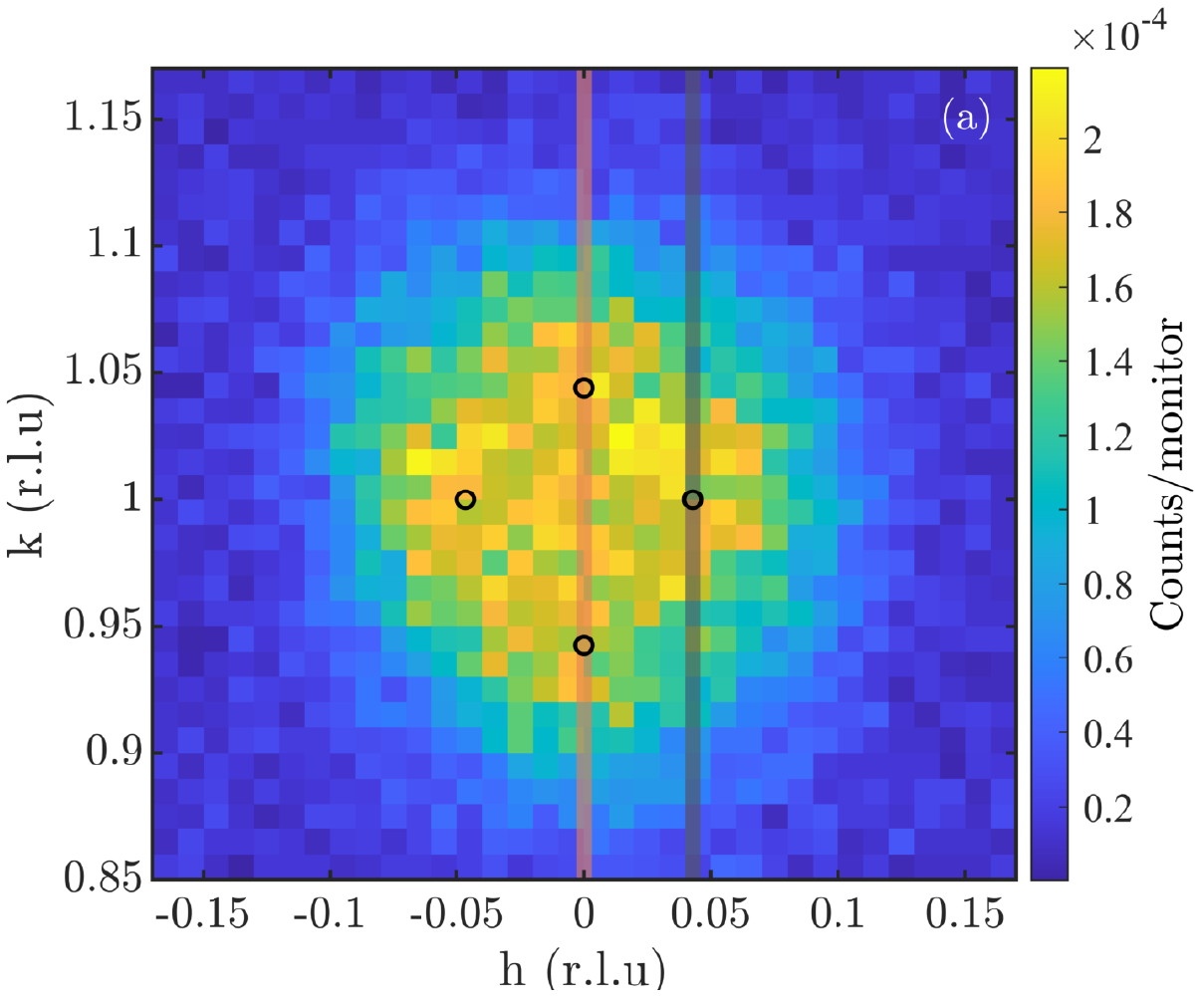}
\includegraphics[width=0.4\textwidth]{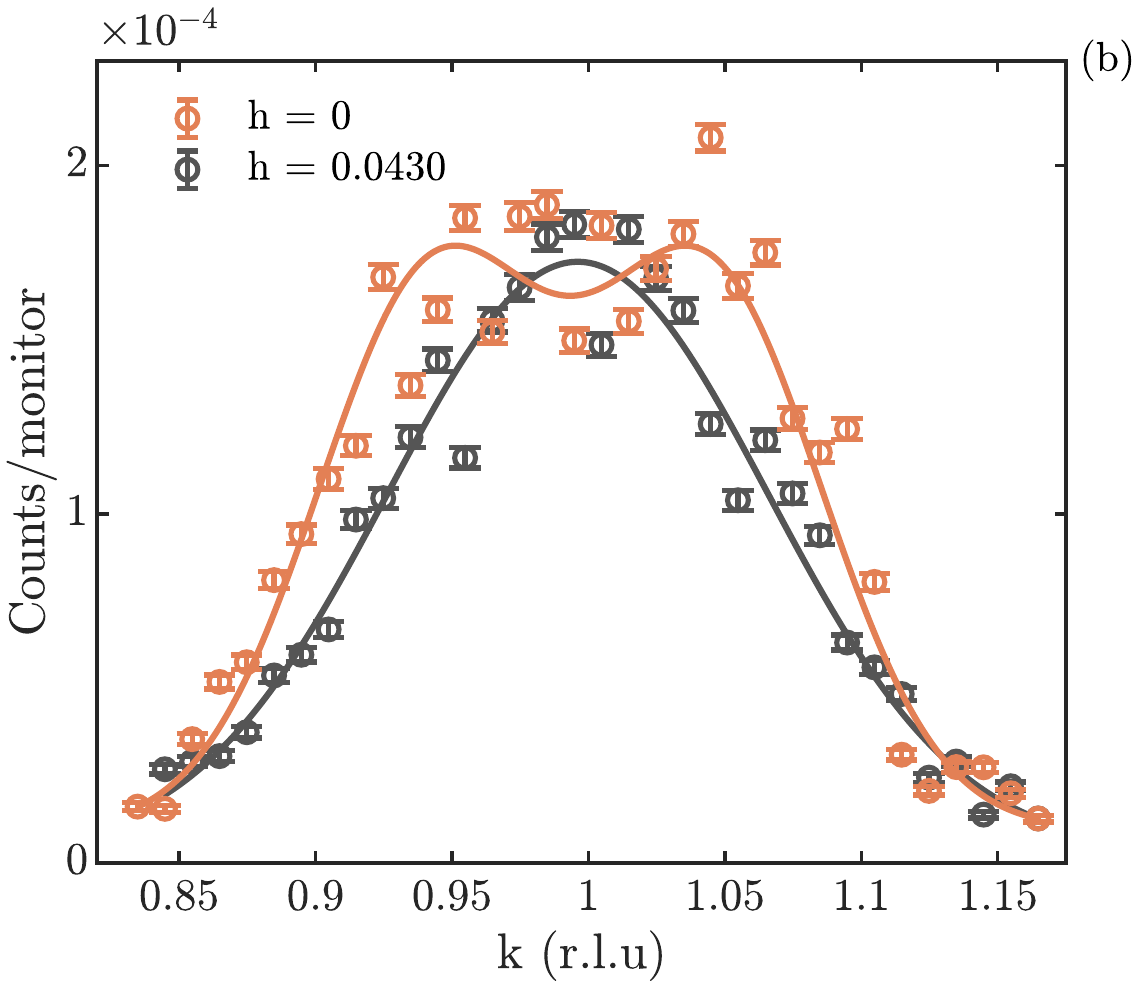}
\caption{\textbf{(a)} Constant energy map of the inelastic magnetic signal with $\hbar \omega = 0.8$~meV measured at $40$~K in the absence of applied magnetic field. The vertical lines show the direction of the two-dimensional scans presented in subfigure (b). The circles indicate centres of the four incommensurate peaks obtained through Gaussian fitting of scans along $(0,k,0)$ and $(h,1,0)$. \textbf{(b)} Two-dimensional cuts though the grid along $(0,k,0)$ (in orange) and $(0.043,k,0)$ (in grey). The solid lines represent the Gaussian fits to the data (single Gaussian in grey and double Gaussian with constrained equal amplitudes and widths in orange).}
\label{fig:LSCO5 grid}
\end{figure}

The onset of superconductivity as a function of doping is also marked, in terms of the magnetic structure, by a transition from a stripe I phase, with charge stripes running along the orthorhombic $a/b$-axis, to a stripe II phase, with charge stripes running along the tetragonal $a/b$-axis. Cartoon pictures of the real space structures, comprising charge rivers and antiferromagnetic spin arrangements, in the two phases are depicted in Figures~\ref{fig:stripes incommensurability}(a) and (c) (note that the local coordinate system is included in each panel). The corresponding neutron patterns, pictured in the inset of Figure~\ref{fig:stripes incommensurability}(b), are $45^\circ$ rotated with respect to one another equivalent to the angle between the tetragonal and orthorhombic coordinate systems. 
In our experiments, we confirm the $45^\circ$ rotation of the stripe signal outside the superconducting dome, i.e. in the case of 5 \% doping, where only weak 2D superconducting correlations have been observed.\\
Figure~\ref{fig:LSCO5 grid}(a) shows a grid scan around the antiferromagnetic $(0,1,0)$ reflection on the LSCO5 sample. We observe a signal which appears elongated along the $h$ and $k$ orthorhombic axis. Note that due to the small incommensurability, resolution limitations prevent the observations of four distinct incommensurate peaks. However, cuts through the grid confirm the  two-peak pattern along $(0, 1\pm\delta,0)$ and not along $(\pm \delta, 1\pm\delta,0)$ (see Figure~\ref{fig:LSCO5 grid}(b)). The data presented in the main paper has been acquired through scans performed along the $(0, 1\pm\delta,0)$ direction.

\newpage

\section{High energy excitations in LSCO5}
We measured the high-energy magnetic excitations in our LSCO5 sample under applied magnetic field on the thermal triple axis spectrometer IN8 at ILL \cite{Hiess2006}. The instrument was configured with a Si(111) monochromator in order to improve the energy resolution and to access a broader incident energy range, since the Si monochromator suppresses background contamination from second order scattered neutrons. Due to spatial constraints in the experimental zone we measured the magnetic signal in the second Brillouin zone around the $(1,2,0)$ reflection. We used a single outgoing wavevector, $k_\textnormal{f} = 4.1$~\AA$^{-1}$, for all the scans performed with energy transfers between $5$ and $50$~meV. The energy resolution (FWHM) for this outgoing wavevector is, however, considerably high ($\sim 6$~meV) meaning that at low energy transfers the signal is contaminated by the tail of the elastic line. In order to minimise the contamination, by reducing the size of the resolution ellipsoid, the lowest energies, $5$ and $7.5$~meV, have also been measured with $k_\textnormal{f} = 2.662$~\AA$^{-1}$.  

\begin{figure}[tb!]
\centering
\includegraphics[width=0.32\textwidth]{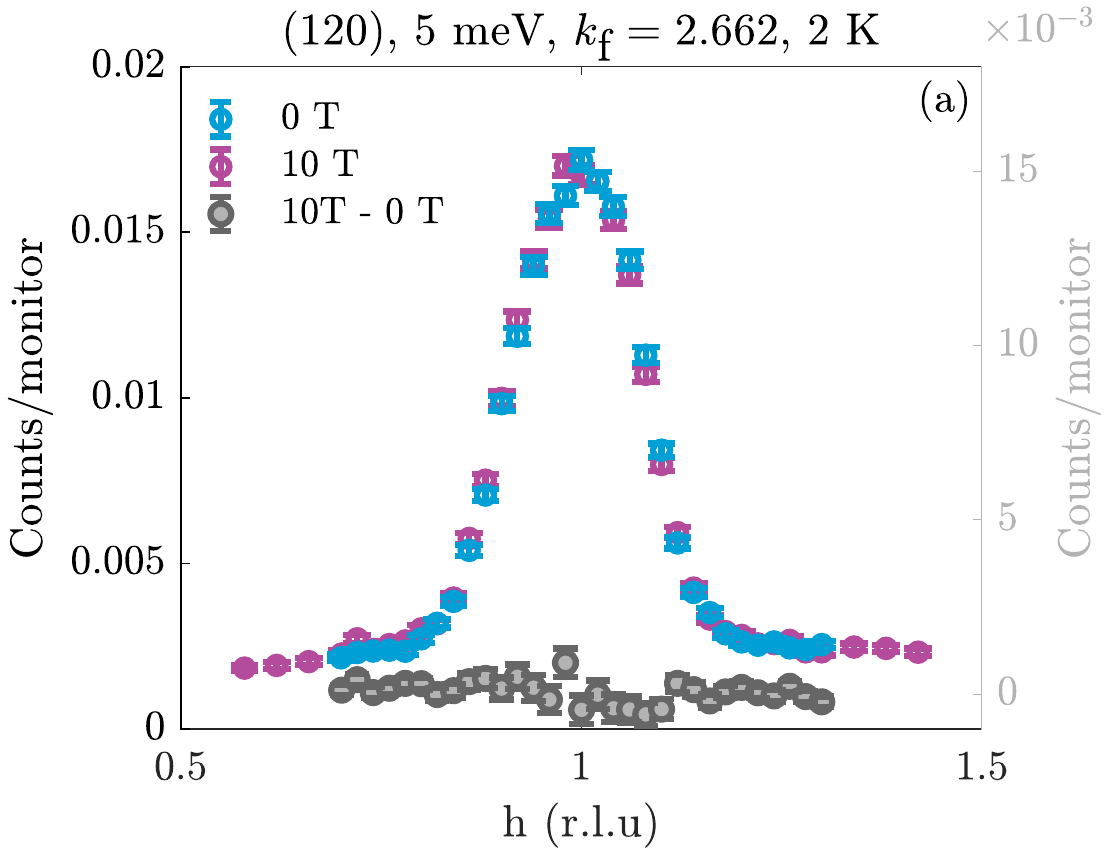}
\includegraphics[width=0.32\textwidth]{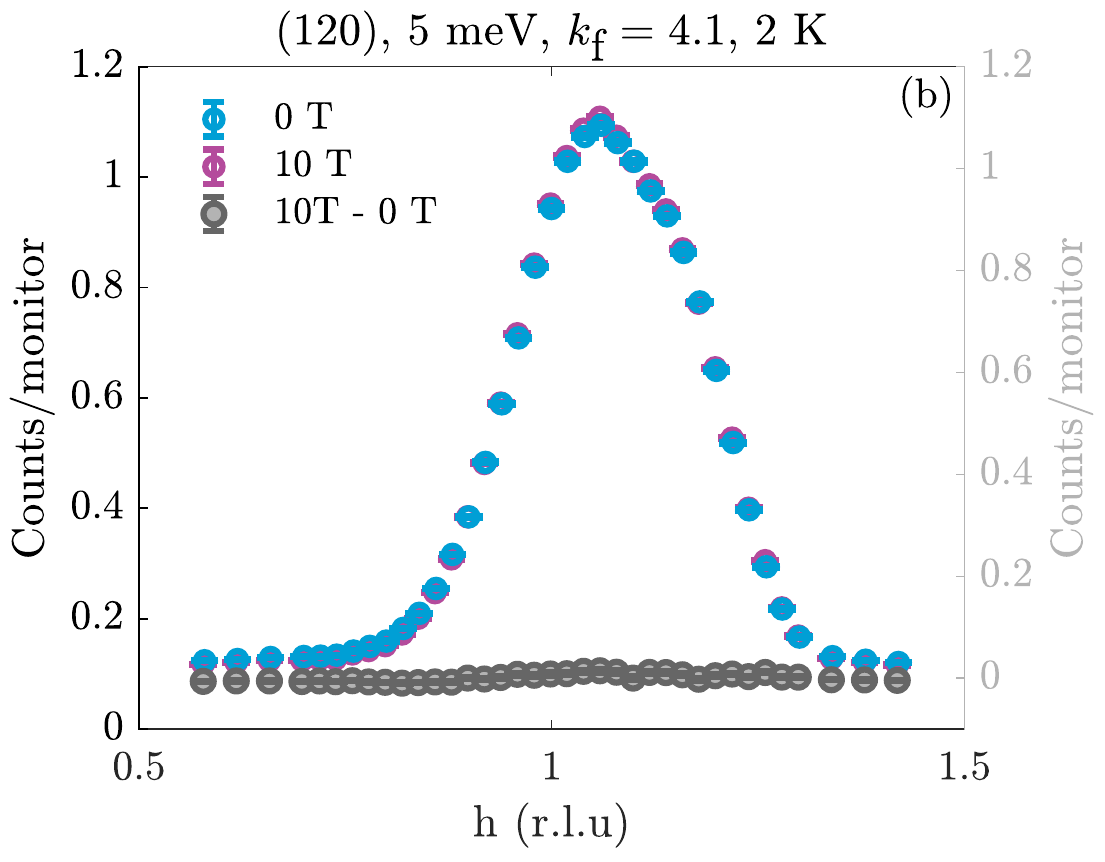}
\includegraphics[width=0.32\textwidth]{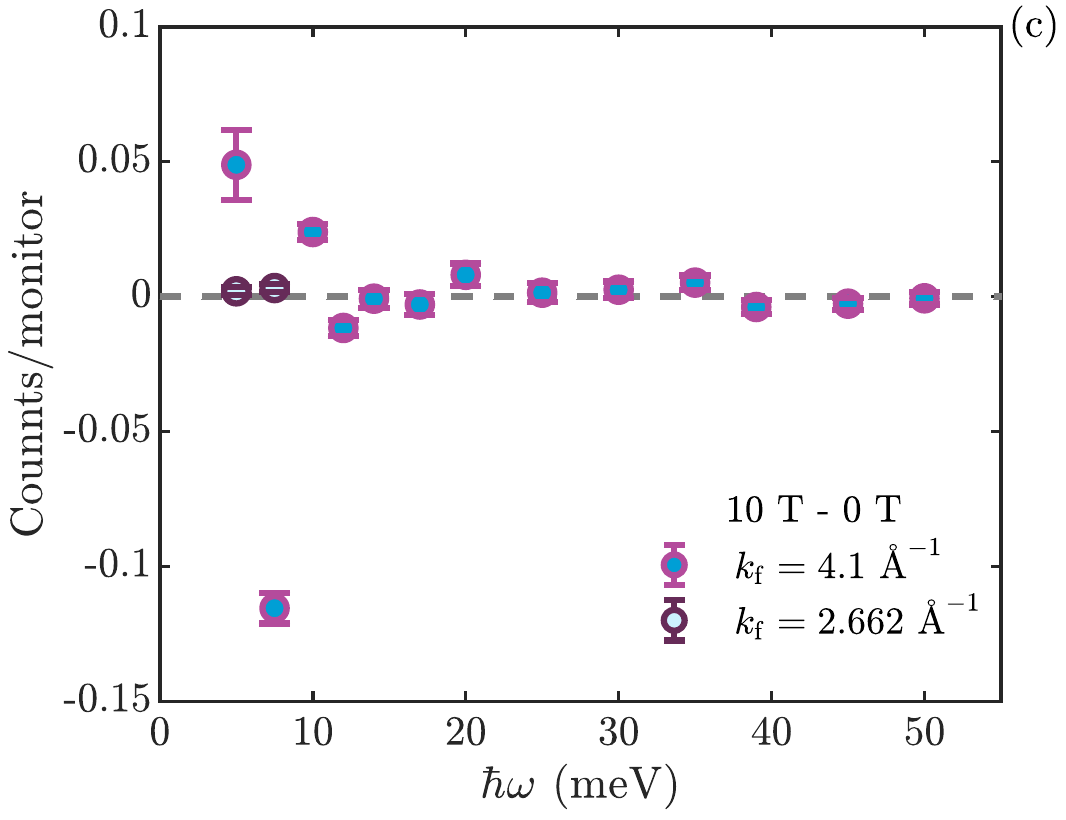}
\caption{\textbf{(a, b)} Inelastic  $5$~meV signal measured at $2$~K as a function of applied magnetic field with two different outgoing wavevectors $k_\textnormal{f}$. In grey a point by point subtraction of the data taken under $10$~T field and in zero applied magnetic field is shown. \textbf{(c)} Variations of the intensity of high-energy fluctuations as a function of applied magnetic field measured at $2$~K. A point by point subtraction of scans taken in $10$~T applied magnetic field and zero field has been employed.}
\label{fig:LSCO5 high energy IN8}
\end{figure}

Additional spurious scattering and a broad $Q$-resolution made the incommensurate peaks appear as a single broad peak (see Figures~\ref{fig:LSCO5 high energy IN8}(a, b)). For this reason the data has not been fitted but rather a point by point subtraction between high field and zero field scans has been used in the analysis. 
Additionally, the acoustic phonon used in the normalisation process to determine the resolution volume could not be measured on the thermal spectrometer. The data is thus presented in counts per monitor rater than absolute units.  
The data acquired below and above the resonance ($E_{\textnormal{cross}} \sim 20$~meV~\cite{fujita2011progress}), shown in Figure~\ref{fig:LSCO5 high energy IN8}(c), confirms that no significant field effect occurs at high energies. %

\section{Temperature dependence of field induced suppression of magnetic fluctuations}
\noindent \underline{LSCO5:}\\
In the main text, we show that in our LSCO5 sample, an applied magnetic field suppresses very low-energy fluctuations ($\hbar\omega<1.5$~meV) at low temperature ($2$~K) while at higher temperature ($40$~K) the field has no effect. At the same time, the only known change in the state of the material taking place between these two temperatures is the onset of magnetic order at $T_\textnormal{N} \sim 15$~K as measured by neutron scattering~\cite{Kofu09} and a diamagnetic response for out-of-plane magnetic fields setting in around $4.2(0.3)$ K. 
\newpage
\noindent \underline{LSCO7} \\
The temperature dependence of the field effect in the case of LSCO7 is depicted in the main text Fig. 4. From this we deduce that the suppression of low-energy fluctuations below $\sim 20$~K approximately correlates with the onset of magnetic order. The raw elastic and inelastic ($\hbar \omega =0.8$ meV) data is shown in Fig. \ref{fig:field temp commparison el inel LSCO7}, supporting the interpretation of a direct, field-induced transfer of spectral weight from low to even lower energies appearing in the static channel.

\begin{figure}[t!]
\centering
\includegraphics[width=0.4\textwidth]{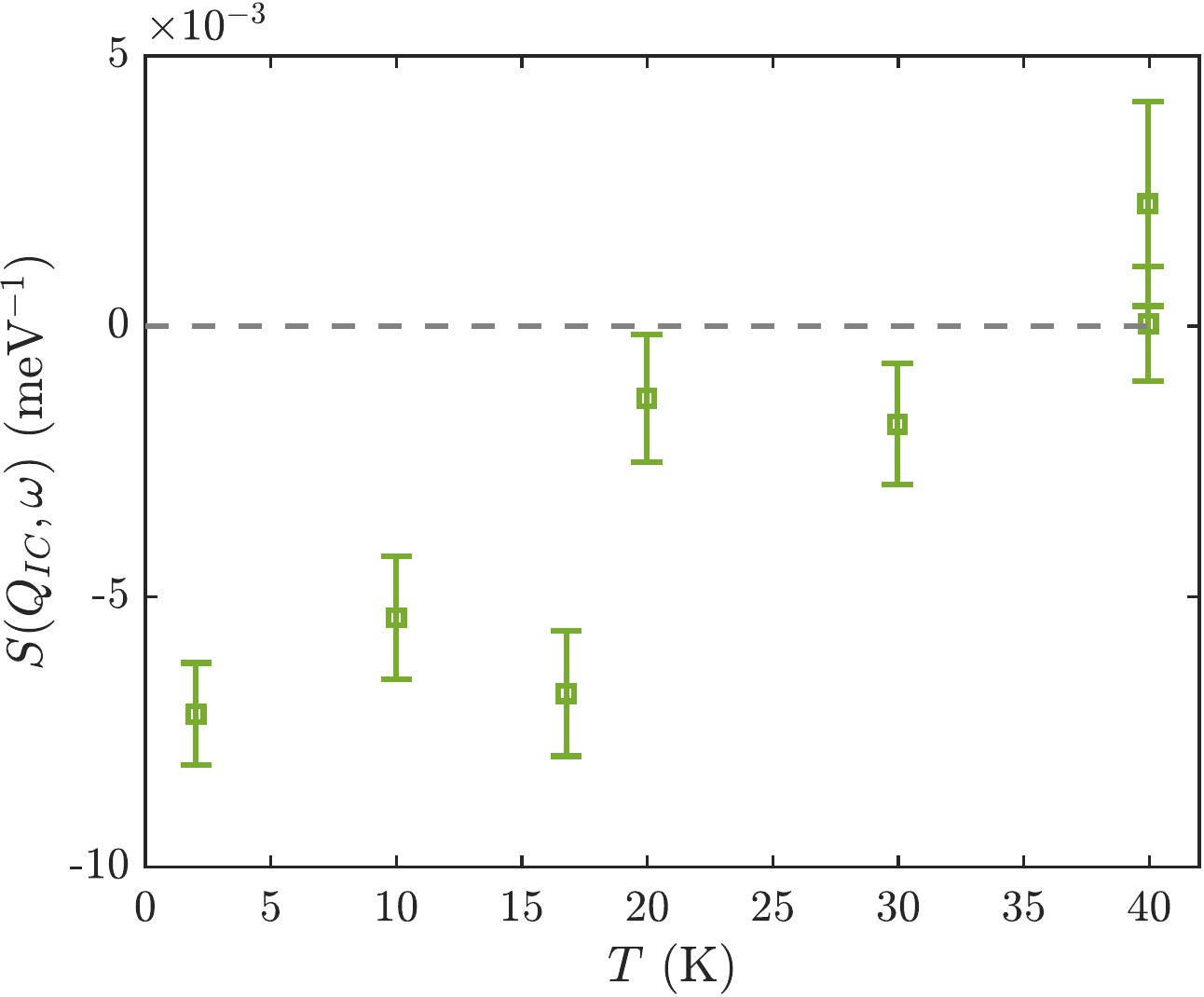}
\includegraphics[width=0.4\textwidth]{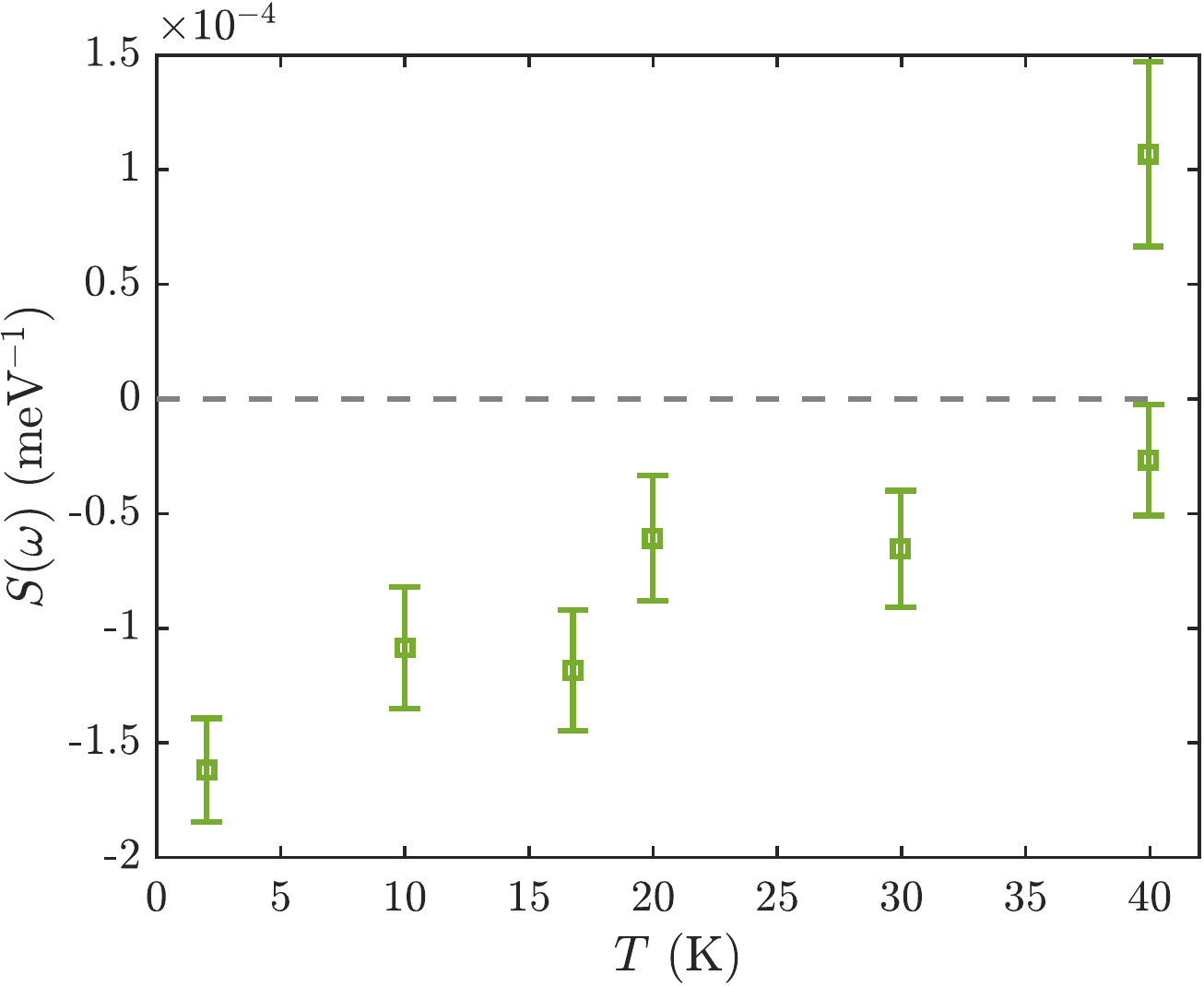}
\caption{Temperature dependence of the field effect of the $ \hbar\omega = 0.8$~meV magnetic fluctuations in the LSCO5 sample. Left shows the amplitude of the peak while right shows the Brillouin Zone averaged value.}
\label{fig:LSCO5 field and temp dependence}
\end{figure}

\begin{figure}[h!]
\centering
\includegraphics[width=0.45\columnwidth]{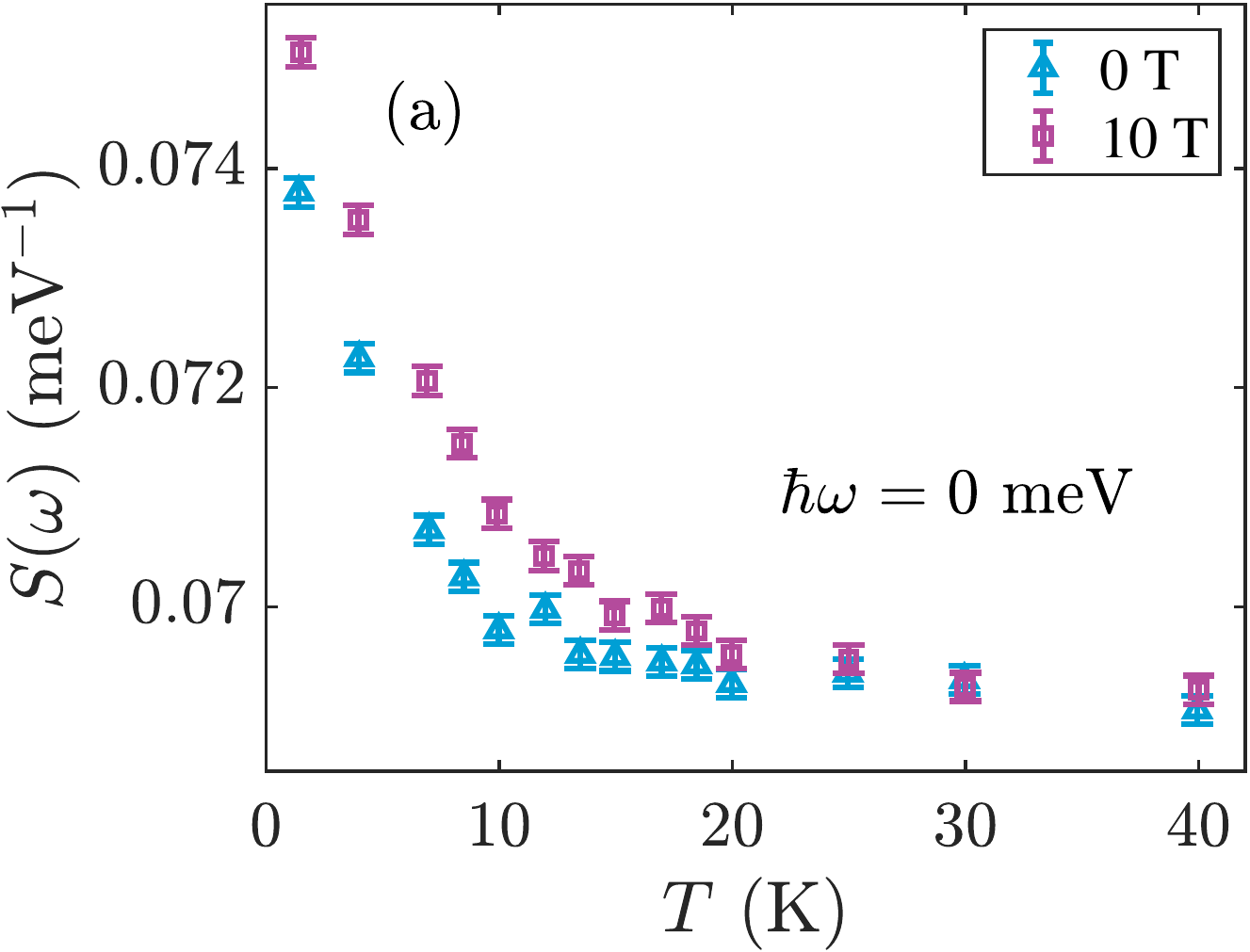}
\includegraphics[width=0.43\columnwidth]{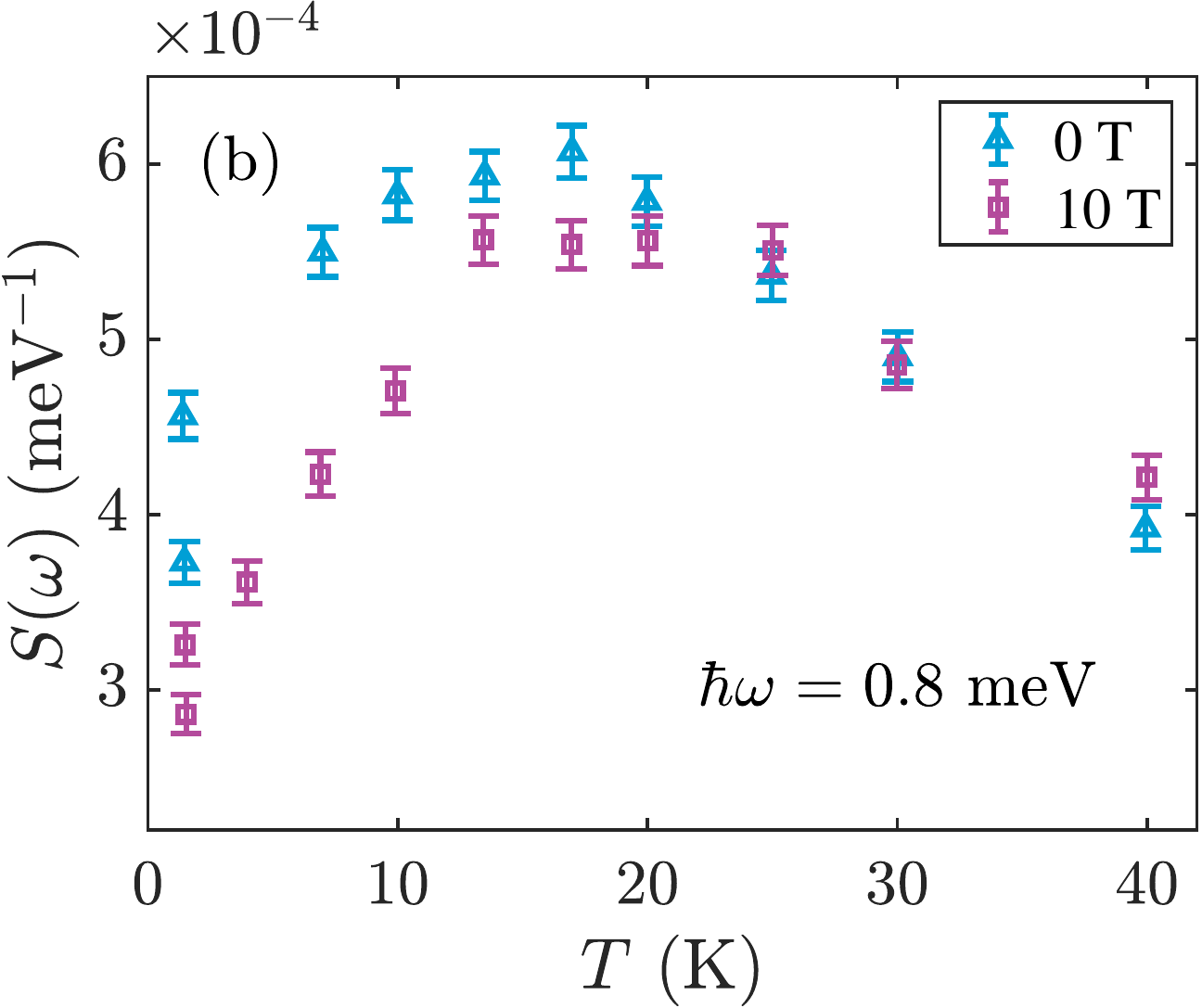}
\caption{Temperature dependence of magnetic order and fluctuations as a function of field in LSCO7. 
\textbf{(a)} Spin correlation $S (\omega=0)$ as a function of temperature in zero magnetic field (blue points) and $H=10$ T (purple points). Data was recorded at the position $Q_\mathrm{IC}=(0.93,0.07,0)$ as 1 point measurements. 
\textbf{(b)} Dynamic spin correlation $S (\omega)$ of $0.8$~meV as a function of temperature in zero magnetic field (blue points) and $H=10$ T (purple points). The data was acquired from fitting of full scans to a Gaussian model as described in the text. }
\label{fig:field temp commparison el inel LSCO7}
\end{figure}

\pagebreak
\typeout{}
\bibliography{LSCO_magnetism_1}